\documentclass[10pt, twocolumn, conference]{IEEEtran}
\IEEEoverridecommandlockouts

\usepackage{subfigure}
\usepackage{paralist}
\usepackage{color}
\usepackage{url}
\usepackage[linesnumbered,ruled,boxed,vlined]{algorithm2e}

\usepackage{algpseudocode}
\usepackage{amsmath}
\usepackage{amssymb}
\usepackage{graphicx}
\usepackage{bm}
\usepackage{paralist}
\usepackage{subfigure}
\usepackage{times}
\usepackage{balance}
\usepackage{cite}
\usepackage{comment}

\newcommand{\separator}{
  \begin{center}
    \rule{\columnwidth}{0.3mm}
  \end{center}
}

\def\ie{{\it i.e.}}

\newcommand{\beq}{\begin{eqnarray*}}
\newcommand{\eeq}{\end{eqnarray*}}
\newcommand{\beqn}{\begin{eqnarray}}
\newcommand{\eeqn}{\end{eqnarray}}
\newcommand{\bemn}{\begin{multiline}}
\newcommand{\eemn}{\end{multiline}}

\renewcommand{\ie
}{{\em i.e., }}

\newcommand\etal{{\em et al.}}

\allowdisplaybreaks
\begin{document}

\title{Permanence based Hidden Community and Graph Recovery in Social Networks 
\thanks{$^*$ Corresponding author. School of Computing, Gachon University, Republic of Korea. (email: jychoi19@gachon.ac.kr) }
              }

%%\author
%%{Jaeyoung Choi, Sangwoo Moon, Jiin Woo, KyunHwan Son, Jinwoo Shin and Yung Yi\\
%%\IEEEauthorblockA{Department of Electrical Engineering\\
%%KAIST, Republic of Korea\\ Emails: \{jychoi14,mununum,jinwoos,yiyung\}@kaist.ac.kr}
%%\thanks{This work was supported by Institute for Information \& communications Technology Promotion (IITP) grant funded by the Korea government (MSIP) (No.B0717-16-0034,Versatile Network System Architecture for Multi-dimensional Diversity). }
%%}

\author
{Jaeyoung Choi$^*$ and Wooseok Sim

%\thanks{This work was supported in part by the National Research Foundation of Korea (NRF) grant funded by the Korea government (MSIP) (No. 2016R1A2A2A05921755) and Institute for Information and communications
%Technology Promotion (IITP) grant funded by the Korea government (MSIP)
%(No.B0717-17-0034, Versatile Network System Architecture for Multidimensional
%Diversity).
%}
}

\maketitle

% \vspace{-1cm}
%\renewcommand{\baselinestretch}{0.99}

\begin{abstract}
Due to the recent development of data analysis techniques, technologies for detecting communities through information expressed in social networks have been developed. Although it has several advantages, including the ability to effectively share recommended items through an estimated community, it may cause personal privacy issues. Therefore, recently, the problem of hiding the real community well is being studied at the same time. As an example, Mittal \etal, proposed an algorithm called NEURAL that can hide this community well based on a metric called permanence. Based on this, in this study, we propose a Reverse NEURAL (R-NEURAL) algorithm that restores the community as well as the original graph structure using permanence.
The proposed algorithm includes a method for well restoring not only the community hidden by the NEURAL algorithm but also the original graph structure modified by recovered edges. We conduct experiments on real-world graphs and found that the proposed algorithm recovers well the hidden community as well as the graph structure.
\end{abstract}
% % A category with the (minimum) three required fields
% \category{H.4}{Information Systems Applications}{Miscellaneous}
% \category{XX}{XX}{XX}
% %\category{F.2.2}{Analysis of Algorithms and Problem Complexity}{Nonnumerical Algorithms and Problems}

% \terms{term1, term2, term3}

% \keywords{keyword1, keyword2, keyword3}

\section{Introduction}
\label{sec:intro}

\begin{figure}[t!]
\begin{center} \centering
\includegraphics[width=1\linewidth]{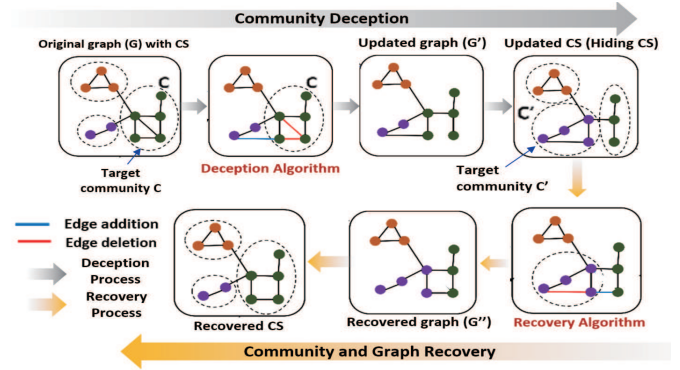}
\caption{Example of community and graph recovery process after the deception of the target community using edge updates. (CS: Community Structure). }
\label{fig:overall}
\end{center}
\vspace{-0.2cm}
\end{figure}

The community detection (CD) problem is a problem that infers which nodes are included in which groups in a network or graph through graph connectivity information. In social networks, it is regarded as an important problem since many users are organized into nodes and each user has complex connections \cite{Papadopoulos2011,Bedi2016}. In this CD problem, the community is inferred by classifying cases in which the connectivity between nodes in a specific group is greater than the connectivity with other groups. Recently, many algorithms have been proposed to solve this CD based on the increase of data in social networks and the development of many machine-learning techniques \cite{Jin2023,Kim2015}. Many of the proposed CD algorithms are made based on the score function for the community structure in the graph. Examples include modularity \cite{Newman2006}, conductance \cite{Leskovec2009}, coverage, or partition quality \cite{Bedi2016}. Modularity is the most used metric which is the ratio between edges in the community and the edges distributed over the graph. This indicates how well it is structured when compared to uniform connections. The conductance is a fraction of edges that have the outside direction from the community and the total number of edges in the community. Coverage is the coverage of all edges in the graph and the edges of the intracommunity.
Partition quality is the ratio of the number of two pairs in the same community.
 In addition, a new score function called permanence has recently been proposed, which is a score function based on the vertices of the graph \cite{Chakraborty2014}. This metric is based on the ratio of connections between inter-community and intra-community. % Details will be explained in Section~\ref{sec:Model}.
Unlike the community detection approaches so far, the authors in \cite{Mittal2021} conducted a study that could well hide the community structure in the graph. They proposed a community deception algorithm named NEURAL, which is a method of selecting a specific edge in the graph based on this permanence score and adding or deleting it to intracommunity or inter-community.
Motivated by the algorithm they proposed, we suggest a method called Reverse NEURAL (R-NEURAL) that can simultaneously restore the original community structure and graph structure after NEURAL (a deception algorithm) is applied as shown in Figure~\ref{fig:overall}. In R-NEURAL, the main approach is to recover edges from hidden graphs based on permanence.

\section{Model and Algorithms}
\label{sec:Model}

\subsection{Preliminaries}
\label{sec:diffusion_model}
We consider an undirected connected graph $G=(V,E)$, where
$V$ is a set of nodes and $E$ is the set of edges
of the form $(i,j)$ for $i, j\in V$. After applying a CD algorithm on $G,$ we assume that there are $k\geq 1$ detected communities and we denote a set of detected communities by $CS=\{C_1, ..., C_k\}.$ We also assume that there are no overlapping communities, \ie $C_i \cap C_j = \emptyset$ for any $i\neq j.$ For each $C_i$, if an edge $(u,v)$ is in $C_i,$ it is denoted by intracommunity edge whereas for an edge  $(u,v)$, where $u \in C_i$ and $v \in C_j$ with $C_i \cap C_j = \emptyset$, we denote it as a intercommunity edge. 

\subsection{Permanence}
As a measure of community structure in the graph, we use the permanence proposed by Chakraborty \etal \cite{Chakraborty2014}. This is a vertex-centric
score function that quantifies how well node $v$ is affiliated with community $C$. To see this, let $I(v)$ be the internal connections of a node $v$ within its own community and let $E_{max} (v)$ be the maximum connections of $v$ to its neighboring communities. Next, we let $C_{in}(v)$ be the fraction of actual and possible number of edges among
the internal neighbors of $v$. Then, the permanence of $v$ in $G$ is given by: 
\begin{align}\label{eqn:permanence}
Perm(v,G) = \frac{I(v)}{E_{max}(v)}\times \frac{1}{deg(v)} - (1-C_{in}(v)),
\end{align}
where $deg(v)$ is the node degree of $v.$ This score function indicates that if the internal connectivity $I(v)$ of vertex $v$ is greater than the external connectivity $E_{max}(v)$, the corresponding value increases and forming a close clique and is divided by the degree of $v$ to normalize it \cite{Chakraborty2014}. As an extension, the permanence of the graph $G$ is defined by $Perm(G)=\sum_{v\in G}Perm(v,G)/|V|,$ 
% \begin{align}\label{eqn:permanence}
% Perm(G)=\frac{1}{|V|}\sum_{v\in G}Perm(v,G),
% \end{align}
where $|V|$ is the number of nodes in $G$. Then the graph permanence is an average of node permanence over the graph. 
\subsection{Community Deception Algorithm - NEURAL \cite{Mittal2021}}
\label{sec:diffusion_model}

 \begin{algorithm}[t!]
 \caption{NEURAL \cite{Mittal2021}}
\label{alg:NEURAL}
{\small

 \KwIn{Graph $G,$ Target community $C,$ Budget $B>0$} 
 \KwOut{Updated graph $G'$}

\smallskip

 $\mathcal{P}_{l,ad} =0$ and $\mathcal{P}_{l,de} =0$;

\While{$B>0$}{
% $\mu(v)\leftarrow 0$\;
\textbf{Step1}: For $u \in C$ and $v \in C'$, where $C'$ is the community of maximum external pull for $u,$ set $G'$ to be the graph with added edge $(u,v)$. Compute $Perm(u,G)$ and $Perm(u,G')$ and choose the best edge satisfying: 

$(u^*,v^*)= \arg\max_{u \in C, v \in C'}Perm(u,G)-Perm(u,G')$;

Set $\mathcal{P}_{l,ad}=\mathcal{P}(G)-\mathcal{P}(G')$ for $(u^*,v^*)$;

\smallskip
\textbf{Step2}: For $w,z\in C$, set $G'$ to be the graph with deleted edge $(w,z)$. Compute $Perm(w,G)$ and $Perm(w,G')$ and choose the best edge satisfying:  

$(w^*,z^*)= \arg\max_{w,z \in C}Perm(w,G)-Perm(w,G')$;

Set $\mathcal{P}_{l,de}=\mathcal{P}(G)-\mathcal{P}(G')$ for $(w^*,z^*)$;
\smallskip

 \If{$\mathcal{P}_{l,ad}\geq \mathcal{P}_{l,de}$ and $\mathcal{P}_{l,ad}>0$}{
   $G\leftarrow (V, E \cup (u^*,v^*))$
   } \ElseIf{ $\mathcal{P}_{l,de}>0$}{
    $G\leftarrow (V, E\setminus(w^*,z^*))$}
    $B \leftarrow B-1$;
    }
    $G'\leftarrow G$;
    
 Return $G'$\;
}
\end{algorithm}
For community deception, the authors \cite{Mittal2021} proposed Network Deception Using Permanence Loss (NEURAL) algorithm in their work. The main idea of the algorithm is that some intracommunity or inter-community edges are updated (added or deleted) based on the permanence difference between before the update and after the edge update. To see this more precisely, we first define a permanence loss \cite{Mittal2021} of two graphs $G$ and $G'$, where $G'$ is the graph result from some edge update on $G,$ as follows:
\begin{align}\label{eqn:loss}
\mathcal{P}_{l}=Perm(G)-Perm(G').
\end{align}
Then, the goal of designing the NEURAL is to maximize the $\mathcal{P}_{l}$. For this, it is necessary to minimize the right term $Perm(G')$ since the permanence of $G$, $Perm(G)$ is fixed if the graph is given. 
The NEURAL algorithm includes two main steps as depicted in Algorithm~\ref{alg:NEURAL}. Step 1 describes the process of finding an edge to add to another community in the target community. That is, 
for given original graph $G$, Target community $C$ and budget $B>0$, which will be used for the edge update, the NEURAL first select nodes pair $(u,v)$, for $u \in C$ and $v \in C'$, where $C'$ is the community of maximum external pull of $u.$ Then, it computes the permanence $Perm(u,G)$ for the original graph $G$ and $Perm(u,G')$ for the updated graph $G'$ with adding the edge $(u,v)$. Next, it computes the difference between them for all $u \in C$ and $v \in C'$ and finds the maximum $(u^*,v^*)$ of the difference (lines 3-4). Then, it calculated the difference of permanences between the original graph $G$ and the updated graph $G'$ as $\mathcal{P}_{l,ad}$. On the other hand, Step 2 includes the process of finding an edge to delete within the same community. In this process, similar to step 1, calculate the permanence of the selected edge to find the edge $(w^*,z^*)$ that maximizes the difference and put it in $\mathcal{P}_{l,de}$ (lines 6-8). \footnote{The reason why it is possible to find edges greedily as in Step 1 and Step 2 is that all permanences satisfy submodularity as shown in \cite{Mittal2021}.
Adding an edge in the inter-community and deleting an edge in the intra-community is an operation that increases the permanence gain as proven in \cite{Mittal2021}, so only these two have been considered.} Finally, the algorithm compares the values of $\mathcal{P}_{l,ad}$ and $\mathcal{P}_{l,de}$ obtained in this way and performs an edge update that satisfies a larger permanence gain, one budget is used (lines 9-13). This process is repeated until the total budget $B$ is exhausted, and the last updated graph is output as $G'$.

% \vspace{-0.1cm}
\subsection{Proposed Algorithm - Reverse NEURAL (R-NEURAL)}
As a simple approach to the recovery of hidden communities that applied the NEURAL algorithm, we propose a Reverse-NEURAL (R-NEURAL) in this paper. This includes the reverse application of the two cases below based on NEURAL. Before describing that, we let $G'$ be the output graph after applying the NEURAL and let $G''$ be the graph after updating edge deleting, or adding on $G'$ to recover the original graph $G$. Then, we introduce permanence gain, $\mathcal{P}_{g}$, as follows: 
\begin{align}\label{eqn:gain}
\mathcal{P}_{g}=Perm(G'')-Perm(G').
\end{align}

The above permanence gain seems quite similar to the permanence loss in \eqref{eqn:loss}. However, the goal of maximizing the permanence loss is to find $G'$ that minimizes the permanence of the new graph $G'$ whereas the goal of permanence gain $\mathcal{P}_{g}$ is to maximize the permanence of the graph $G''$ recovered from $G'$. To do this, we reverse the following two in Algorithm~\ref{alg:NEURAL}, which is described in Algorithm~\ref{alg:R-NEURAL}.
\smallskip
\begin{itemize}
\item [$(1)$] In Step 1, since inter-community edges are newly connected in Step 1 of the NEURAL algorithm, we work to find and remove edges that exist between intercommunity. Due to the lack of knowledge of which edges have been added, R-NEURAL uses also a similar measure, permanence gain $\mathcal{P}_{g,de}$, to select edges to be deleted.
\smallskip

\item [$(2)$] In Step 2, based on the method of finding the permanence loss while removing the intracommunity edge in NEURAL, we consider the addition of an intracommunity edge to $G'$ and raise the permanence of $G''$ to $\mathcal{P}_{g,ad}$.

\end{itemize}
Finally, we compare $\mathcal{P}_{g,de}$ and $\mathcal{P}_{g,ad}$ and proceed with the operation for the larger one and consume one budget. We examine how well the community structure hidden by NEURAL can be recovered through this simple opposite approach.

% \smallskip
% \begin{definition}(Submodular Function)
%   \label{prop:simple_prob}
% A function $f$ is called submodular if $$f(S\cup x)-f(S) \leq f(T\cup x)-f(T)$$ for any subset $S \subseteq T$ where $x \notin T$ .
% \end{definition}

\begin{algorithm}[t!]
 \caption{Reverse NEURAL (R-NEURAL)}
\label{alg:R-NEURAL}
{\small

 \KwIn{Output Graph $G',$ Target community $C,$ Budget $B>0$} 
 \KwOut{Updated graph $G''$}

\smallskip

 $\mathcal{P}_{l,ad} =0$ and $\mathcal{P}_{l,de} =0$;

\While{$B>0$}{
% $\mu(v)\leftarrow 0$\;
\textbf{Step1}: For $u \in C$ and $v \in C'$, where $C'$ is the community of maximum external pull for $u,$ set $G''$ to be the graph with deleted edge $(u,v)$.  Compute $Perm(u,G)$ and $Perm(u,G')$ and choose the best edge $(u^*,v^*)$ with 

\quad $\arg\max_{u \in C, v \in C'}Perm(u,G'')-Perm(u,G')$;

Set $\mathcal{P}_{g,de}=\mathcal{P}(G'')-\mathcal{P}(G')$ for $(u^*,v^*)$;

\smallskip
\textbf{Step2}: For $w,z\in C$, set $G''$ to be the graph with added edge $(w,z)$. Compute $Perm(w,G)$ and $Perm(w,G')$ and choose the best edge satisfying:  

$(w^*,z^*)= \arg\max_{w,z \in C}Perm(w,G'')-Perm(w,G')$;

Set $\mathcal{P}_{g,ad}=\mathcal{P}(G'')-\mathcal{P}(G')$ for $(w^*,z^*)$;
\smallskip

 \If{$\mathcal{P}_{g,de}\geq \mathcal{P}_{g,ad}$ and $\mathcal{P}_{g,de}>0$}{
   $G'\leftarrow (V, E\setminus(u^*,v^*))$
   } \ElseIf{ $\mathcal{P}_{g,ad}>0$}{
    $G'\leftarrow (V, E\cup (w^*,z^*))$}
    $B \leftarrow B-1$;
    }
    $G''\leftarrow G'$;
    
 Return $G''$\;
}
\end{algorithm}

\smallskip

\section{Numerical and Simulation Results}\label{sec:simulation}
\begin{figure}[t!]
\begin{center} \centering
\includegraphics[width=0.7\linewidth]{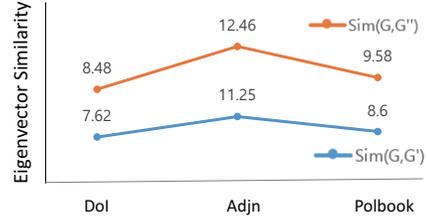}
\caption{Eigenvector similarity from the original graph $G$. The orange line shows how close the graph $G''$ is recovered to $G$ compared to $G'$.}
\label{fig:sim}
\end{center}
\vspace{-0.2cm}
\end{figure}

In the simulation, we apply several different CD algorithms for $(i)$ original graph $G$, $(ii)$ updated graph $G'$ after NEURAL, and $(iii)$ recovered graph $G''$ after R-NEURAL, respectively. As an original graph, we use three different real-world graphs: Dolphin social network (Dol, 62 nodes) \cite{graph}, and word adjacencies (Adjn, 112 nodes) \cite{graph}, and Polbook (105 nodes) \cite{graph}, respectively. As evaluation metrics, we consider three measures: modularity (M), coverage (C), and partition quality (PQ). In the experiments, we set $B=0.3|V_C|$, \ie 30\% of the size of the target community $C$ as in \cite{Mittal2021}.
As result, in Table~\ref{graph}, we obtain three evaluation results for two real-world graphs, we see that there is an increase in most evaluation metrics for each graph $G''$ compared to those of $G'$. Among them, both graphs show that the partition quality score improves the most, which is due to the process of community recovery based on permanence.
Next, we use two CD algorithms: Louvain \cite{Blondel2008} and InfoMap \cite{Rosvall2008} in Table~\ref{CDA}. We obtain the three evaluation results with respect to the applied CD algorithms for one real-world graph (Dol). In this case, we also check that the detection performance is improved in the recovered graph $G''$ in most cases. 
 Finally, to check how similar the recovered graph was to the original graph, we obtained the similarity measure between the two graphs using the well-known eigenvector similarity. To do this, we first measured the similarity between $G$ and the recovered graph $G''$ (sim($G$,$G''$)) together with the similarity between the original graph $G$ and the updated graph $G'$ (sim($G$,$G'$)) for deception in Figure~\ref{fig:sim}. We compared the similarity of Dol, Adjn, and Polbook graphs. As a result, all graph $G''$ in the three real-world graphs can recover the original graph from the updated graph $G'$.

\begin{table}[t!]
\renewcommand{\arraystretch}{1}
\caption{Two real-world graphs: Dol (left) and  Adjn (right).} \label{graph} \centering \begin{tabular}{c||c|c|c||c|c|c}
\hline  Metrics  & $G$ (Dol)   & $G'$  & $G''$ & $G$ (Adjn)   & $G'$  & $G''$ \\
\hline M & 0.5202 & 0.4822 & 0.4866 & 0.2941 & 0.2911 & 0.2920\\
\hline  C & 0.7447 &0.7119 &0.7123 & 0.4682 & 0.4479 & 0.4448\\
\hline  PQ & 0.8381 &0.8239 &0.8324 & 0.8394 & 0.8323 & 0.8351\\
\hline
\end{tabular} \end{table}
\vspace{-0.2cm}
\begin{table}[t!]
\renewcommand{\arraystretch}{1}
\caption{Two CD algorithms: Louvain (left) and Infomap (right).} \label{CDA} \centering \begin{tabular}{c||c|c|c||c|c|c}
\hline  Metrics  & $G$ (Dol)  & $G'$  & $G''$ & $G$ (Dol)   & $G'$  & $G''$ \\
\hline M & 0.5205 & 0.4796 & 0.4854 & 0.5219 & 0.4198 & 0.4570\\
\hline  C & 0.7418 &0.7068 &0.7063 & 0.7494 & 0.8400 & 0.7198\\
\hline  PQ & 0.8407 &0.8285 &0.8525 & 0.8358 & 0.7298 & 0.8371\\
\hline
\end{tabular} \end{table}

% \begin{table}[t!]
% \renewcommand{\arraystretch}{1}
% \caption{Graph Similarities.} \label{para_1} \centering \begin{tabular}{c||c|c|c}
% \hline  Graphs Similarity &  DoI  &  Adjn &  Polbooks    \\
% \hline Eigenvector$(G,G')$ & 7.6212 & 11.2510 & 8.5959 \\
% \hline  Eigenvector$(G,G'')$  & 8.4757 &12.4595 &9.5764 \\
% \hline
% \end{tabular} \end{table}

\section{Conclusion}
\label{sec:conclusion}
In this study, we proposed a reverse NEURAL algorithm that restores the community and graph structure using permanence.
From the simulation on real-world graphs, we have checked that our proposed algorithm recovered not only the target community hidden by the NEURAL algorithm but also the original graph structure modified by edges. As future work, we will consider an additional query process for the selected edges in the algorithm to check whether it was in the original graph or not to improve the recovery performance. 

\balance
{
\renewcommand{\baselinestretch}{1}
\bibliographystyle{IEEEtran}

}

\end{document}